# Machine-Learning Interatomic Potentials Enable First-Principles Multiscale Modeling of Lattice Thermal Conductivity in Graphene/Borophene Heterostructures

Bohayra Mortazavi*[a,b], Evgeny V. Podryabinkin[c], Stephan Roche[d,f],

Timon Rabczuk[g], Xiaoying Zhuang[a,g]** and Alexander V. Shapeev[c]

[a]*Chair of Computational Science and Simulation Technology, Department of Mathematics and Physics, Leibniz Universität Hannover, Appelstraße 11, 30157 Hannover, Germany.*
[b]*Cluster of Excellence PhoenixD (Photonics, Optics, and Engineering–Innovation Across Disciplines), Gottfried Wilhelm Leibniz Universität Hannover, Hannover, Germany.*
[c]*Skolkovo Institute of Science and Technology, Skolkovo Innovation Center, Nobel St. 3, Moscow 143026, Russia.*
[d]*Catalan Institute of Nanoscience and Nanotechnology (ICN2), CSIC and BIST, Campus UAB, Bellaterra, 08193 Barcelona, Spain.*
[f]*ICREA Institució Catalana de Recerca i Estudis Avancats, 08010 Barcelona, Spain*
[g]*College of Civil Engineering, Department of Geotechnical Engineering, Tongji University, Shanghai, China.*

## Abstract

One of the ultimate goals of computational modeling in condensed matter is to be able to accurately compute materials properties with minimal empirical information. First-principles approaches such as the density functional theory (DFT) provide the best possible accuracy on electronic properties but they are limited to systems up to a few hundreds, or at most thousands of atoms. On the other hand, classical molecular dynamics (CMD) simulations and finite element method (FEM) are extensively employed to study larger and more realistic systems, but conversely depend on empirical information. Here, we show that machine-learning interatomic potentials (MLIPs) trained over short ab-initio molecular dynamics trajectories enable first-principles multiscale modeling, in which DFT simulations can be hierarchically bridged to efficiently simulate macroscopic structures. As a case study, we analyze the lattice thermal conductivity of coplanar graphene/borophene heterostructures, recently synthesized experimentally (*Sci. Adv.* 2019; **5**: eaax6444), for which no viable classical modeling alternative is presently available. Our MLIP-based approach can efficiently predict the lattice thermal conductivity of graphene and borophene pristine phases, the thermal conductance of complex graphene/borophene interfaces and subsequently enable the study of effective thermal transport along the heterostructures at continuum level. This work highlights that MLIPs can be effectively and conveniently employed to enable first-principles multiscale modeling via hierarchical employment of DFT/CMD/FEM simulations, thus expanding the capability for computational design of novel nanostructures.

Corresponding authors: *bohayra.mortazavi@gmail.com; **zhuang@ikm.uni-hannover.de

From the engineering point of view, numerical modeling is currently a fundamental aspect of structural design, which not only substantially reduces the final costs of a product but also enables the optimization toward the improved performance. However, before conducting an engineering simulation, materials properties are ought to be evaluated accurately. In comparison with conventional materials, experimental techniques for the characterization of nanomaterials properties are substantially more complicated, time-consuming and expensive as well. More importantly, for nanomaterials the experimentally reported properties may show considerable scatterings, stemming from diverse sources of uncertainties in the measurements. Like other engineering products, for the practical application of nanomaterials in various technologies, developments of accurate modeling approaches are critical to facilitate the design and further optimizations. In recent years theoretical simulations have played a major role in the astonishing advances in the field of materials science. In this regard, modeling enables researchers to examine the stability and explore the properties of novel materials and structures purely through computer simulations. Notably, first-principles simulations can already be employed to find possible synthesis routes for the design of novel materials[1–3]. As a recent example, boron nanosheets with different atomic lattices were epitaxially grown over the silver surface [4,5], a fabrication process that was originally proposed by the density functional theory (DFT) simulations[6,7].

The main drawback of first-principles DFT calculations is nonetheless related to their demanding computational cost, which limits the maximal size of studied systems to only several hundreds, or at most a few thousand atoms. Moreover, the computational costs of common DFT simulations normally scale exponentially with the number of atoms, which jeopardize the numerical exploration of large and disordered material models such as the amorphous graphene [8]. Classical molecular dynamics (CMD) simulation is also among the most popular numerical approaches, and has been extensively employed to explore the properties of complex nanostructured materials. Unlike DFT simulations, the computational cost of CMD calculations scales linearly with the number of atoms, giving access to million-atom scale modeling. However, the accuracy of CMD results strongly depends on the precision of the interatomic potentials in describing energies and forces. As a well-known example, despite the rather simple bonding mechanism in the planar full-$sp^2$ carbon system, most of the currently available interatomic potentials cannot accurately reproduce the thermal conductivity of graphene. Additionally, for novel materials and structures, it is a challenging task to find an interatomic potential that maintains structural stability, irrespective of the accuracy in estimating the basic mechanical or vibrational properties. It is clear that in comparison with the DFT counterpart, the computational advantage of MD simulations comes with the costs of declined accuracy. On the other side, continuum mechanic based method like the finite element method (FEM) offer robust solutions to study practical engineering problems, but in these methods, the properties of the materials should be fully known prior to launching a simulation. It is thus conspicuous that for studying the properties and responses of nanomaterials, the development of multiscale approaches, solving each method's drawback, is crucially needed.

The latest advances in the field of machine-learning methods have offered novel solutions to address critical challenges for a number of problems, especially in materials science [9–13]. For example, as discussed in numerous studies [14–20] machine-learning techniques are expected to revolutionize the materials discovery and design. One of the latest advances in this regard, is the use of machine-learning interatomic potentials (MLIPs) to substantially enhance the accuracy of CMD simulations. Recently, MLIPs have been successfully employed to predict novel materials [21,22] and examine lattice dynamics [23,24] and thermal conductivity [25,26]. As proven in numerous recent studies [23,25,27], MLIPs enable CMD simulations to be conducted within the DFT level accuracy for the computed energies and forces, but with computational costs scaling linearly with the number of atoms. Another remarkable advantage of MLIPs is that being derived from DFT simulations, they can be trained for a specific material composition and are thus less affected by the flexibility issue of standard CMD method. Accordingly, MLIPs offer unprecedented possibility to marry first-principles accuracy with multiscale modeling. To illustrate such strategy, here we examine the lattice thermal conductivity of graphene/borophene heterostructures[28], as a truly challenging system to simulate accurately with conventional methods. To date, there is no available classical interatomic potential that can accurately reproduce the structural properties of borophene and borophene/graphene nanosheets. Moreover, for a well-studied system like graphene, while the majority of interatomic potentials provide structural and elastic constants with a sufficient accuracy, when applied to the calculation of the lattice thermal conductivity, variation of one order of magnitude is observed. For the case of graphene, the experimentally measured thermal conductivities lie in the range 1500-5300 W/mK[29–32], while CMD based estimates by the original Tersoff [33], AIREBO [34], REBO [35] and optimized Tersoff [36] give values of 870[37], 709[38], 350[39] and ~3000 W/m.K[40,41], respectively. Accordingly it is evident that the prediction of lattice thermal conductivity using classical interatomic potential remains a highly challenging task. In addition, when using CMD simulations to evaluate the interfacial thermal conductance, the interatomic potential must exhibit both high stability and accuracy, otherwise the calculations fail to simulate the steady-state heat transfer[42–44]. Therefore the stability of simulations is a critical issue for the modeling of graphene/borophene structures, not only because of their different lattices but also due to the possibility of formation of diverse types of defects at their interface.

The main steps within the first-principles hierarchical multiscale modeling framework proposed here are summarized in Fig. 1. This includes four key steps: (1) DFT simulations; (2) development of MLIPs; (3) CMD simulations and (4) FEM modeling of effective lattice thermal conductivity. Within the DFT step, we first conduct the energy minimization of graphene and borophene lattices. Next, ten different possible grain boundaries (GBs) between the graphene and borophene lattices (find Fig. 2a) are studied. Since we conduct the DFT simulations within the plane-wave approach, the constructed models are periodic in planar directions, so that it is possible to construct two different grain boundaries in every DFT interface model (see Section 2 in the Supplementary Information document). To create the required training sets for the development of MLIPs, ab-initio molecular dynamics (AIMD) simulations are performed. These simulations are carried out for pristine phases (pure graphene or borophene) and heterostructures with geometry optimized interfaces at different temperatures of 100, 300, 600 and 700 K, each one with less than 1000 time steps. Since the AIMD trajectories are correlated within short time periods, only every 10$^{th}$ steps of the original trajectories are included in the training sets. Next, moment tensor potentials (MTPs)[45] are parameterized to describe the interatomic interactions. Similarly to classical counterparts, MTPs also include parameters which are optimized over the training configurations provided by the AIMD simulations. In this work, two types of MTPs are developed, mono-elemental potentials to simulate the pristine graphene or borophene and binary potentials for the heterostructure samples. In the latter case, the created training sets not only include the AIMD trajectories from the constructed heterostructures but also those from the pristine graphene and borophene lattices. For the computational efficiency, MTPs are first trained over subsampled AIMD trajectories. After the preliminary training of MTPs, the accuracy of the trained potentials is evaluated over the full AIMD trajectories and the configurations with high extrapolations grades[27] are identified. Such selected configurations are then added to the original training sets and the final MTPs developed by retraining the updated clean potentials over the updated training sets (see Section 1 in the Supplementary Information document). After the MTPs are trained, they are used in the third step to evaluate the thermal conductivity of pristine phases or calculate the interfacial thermal conductance of grain boundaries via CMD simulations. In the last step, the effective lattice thermal conductivity is evaluated with the FEM method, using the input data provided by the third step.

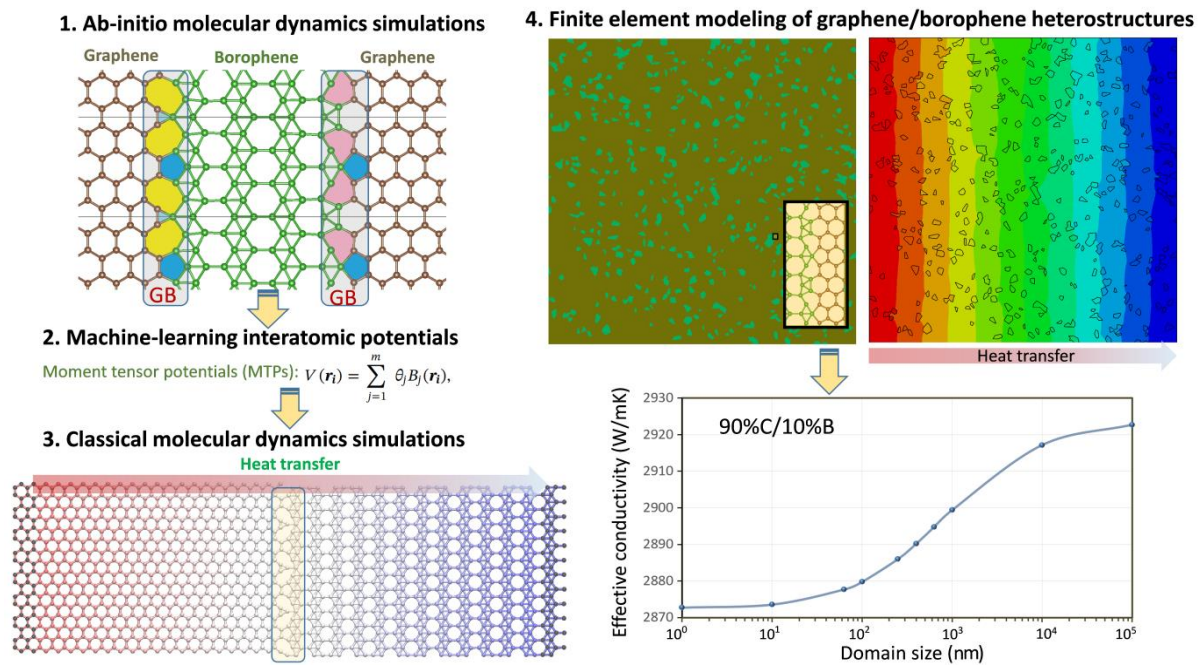

**Fig. 1**, Main steps of the proposed first-principles multiscale modeling framework to simulate the lattice thermal conductivity of graphene/borophene heterostructures.

In Fig. 2a, the atomic configurations of ten different graphene/borophene grain boundaries constructed in this study are illustrated. In this work, six different graphene and borophene heterostructure models are constructed, and from those models ten different grain boundaries are selected for the CMD simulations (find Fig. S1). We remind that from a theoretical point of view, the grain boundaries in graphene are made from a series of pentagon/heptagon pairs[46]. For the case of $MoS_2$, which also shows the hexagonal unit cell, according to high-resolution electron microscopy results, grain boundaries however contain diverse forms of pentagon-heptagon (5-7), tetragon-tetragon (4-4), tetragon-hexagon (4-6), tetragon-octagon (4-8) and hexagon-octagon (6-8) rings[47–49]. Because of the different atomic lattices of borophene and graphene and depending on the various tilting angles, graphene/borophene grain boundaries show diverse configurations. From the constructed grain boundaries shown in Fig. 2a it is clear that they mainly include tetragon, pentagon, hexagon and octagon dislocations, but nonagon rings may also form as found in the case of GB-1. Non-equilibrium molecular dynamics (NEMD) simulations are then performed to access the interfacial thermal conductance and lattice thermal conductivity of pristine borophene. To this end we used the LAMMPS [50] package along with the trained MTPs to introduce the atomic interactions. In Table S1, we examine the accuracy of the trained MTPs for the graphene/borophene interfaces over additional 4 ps of AIMD trajectories at 300 K. Our analysis over the AIMD testing data reveals that the errors in the absolute energy of the systems are in the order of a few meV, confirming the high accuracy of the trained MTPs.  In the NEMD approach periodic boundary conditions are applied along the planar directions using a simulation time step of 0.5 fs. As shown schematically in Fig. 2b, to simulate the steady-state heat transfer, we first relax the structures at room temperature using the Nosé-Hoover thermostat (NVT) method. Then a few rows of atoms at the two ends were fixed and the rest of the simulation box is divided into 22 slabs. Next a temperature difference of 20 K is applied between the first (hot) and last (cold) slabs. In this process, the desired temperatures at the two ends are controlled by the NVT method, while the remaining of the system is simulated without applying a thermostat. As shown in Fig. 2c for a sample of grain boundaries, to keep the applied temperature difference at every simulation time step an amount of energy is added to the hot slab and another amount of energy is removed from the cold slab by the NVT thermostat. As can be seen from Fig. 2c, the amounts of the energy added and removed to the system remain under control (that show linear patterns), confirming that the system stays under steady-state heat transfer condition. The slope of these energy curves can be used to calculate the applied steady-state heat flux ($H_f$). As shown in Fig. 2d, due to the existence of grain boundary, the temperature profile exhibits a sudden change at the interface (ΔT). It is noticeable that temperature gradient within the graphene region is negligible as compared with the borophene section, suggesting a considerably higher lattice thermal conductivity of graphene. The grain boundary thermal conductance can be calculated as $H_f$/ΔT. For the pristine borophene, the temperature profile however illustrates a constant gradient, which can be used to estimate the thermal conductivity using the one-dimensional form of the Fourier law. In Fig. 2e, the calculated interfacial thermal conductance for the considered grain boundaries are compared. Notably, the thermal conductances of different grain boundaries are close. We also examine the length dependence and found that it does not affect the estimated thermal conductance, in agreement with a recent study on polycrystalline $MoS_2$[51] and graphene/h-BN heterostructures[52]. These observations reveal that the interfacial thermal resistance mainly stems from the very different phonon dispersion relations of graphene and borophene (find Fig. S2), in contrast with those of polycrystalline sheets in which the misorientation angle of adjacent sheets and density and type of dislocations cores play the critical role [51,52]. Since the thermal conductance does not show substantial dependence on the geometries of the formed defects at graphene/borophene interfaces, it is thus expected that including more extensive grain boundary configurations should not lead to considerable changes in the estimated effective lattice thermal conductivity of heterostructures.

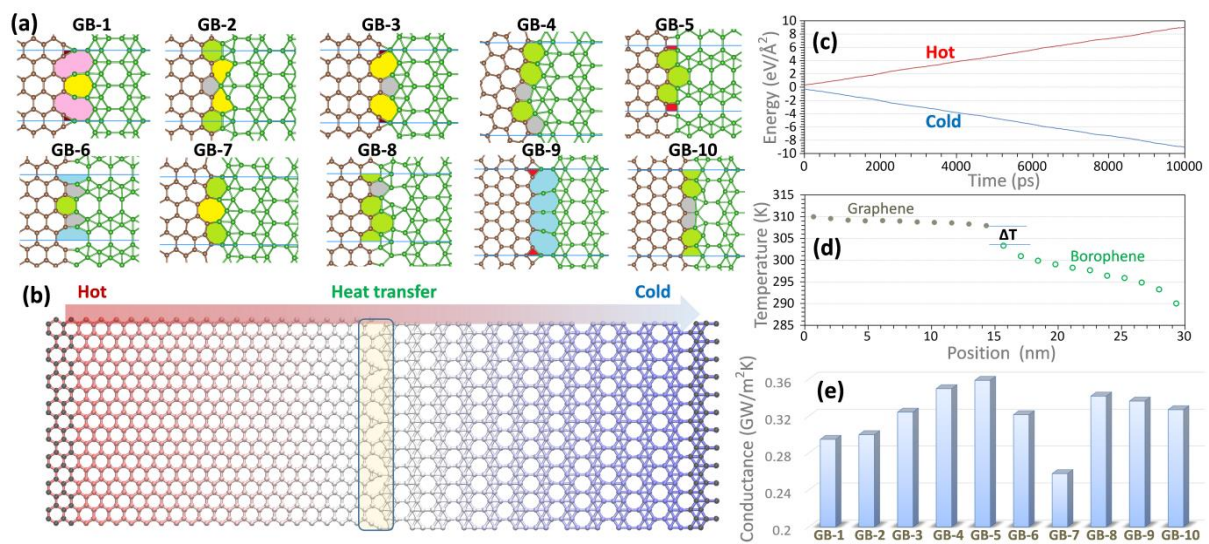

Fig. 2, (a) Atomic configurations of constructed graphene/borophene grain boundaries (GB), (b) schematic illustration of non-equilibrium molecular dynamics (NEMD) method, (c) energy values added to the hot slab and removed from the cold slab by the NVT thermostat during every simulation time step, (d) established temperature profile showing a sudden drop at the interface (e) estimated interfacial thermal conductance of considered grain boundaries in panel (a).

The length effect on the NEMD predictions for the lattice thermal conductivity of borophene monolayer at room temperature is plotted in Fig. 3a. Unlike the graphene, borophene shows anisotropic transport properties and therefore the calculations are conducted along the armchair and zigzag directions. Sharp initial increases in the predicted lattice thermal conductivities by increasing the sample length are observable. This length effect on the thermal conductivities suppresses at higher lengths and finally converges and reaches the diffusive heat transfer regime. As a common approach, the thermal conductivity of borophene at infinite length, $k_\infty$, can be calculated by an extrapolation of the NEMD results for the samples with finite lengths, $k_L$, using the first-order rational curve fitting via $1/k_L=(1+\Lambda/L)/k_\infty$[53,54], where $\Lambda$ is the effective phonon mean free path. By assuming the thickness of 2.9 Å, the diffusive lattice thermal conductivity of borophene at room temperature along with the armchair and zigzag directions were estimated to be 52 and 112 W/mK, respectively.

Another alternative to calculate the lattice thermal conductivity is to solve the Boltzmann transport equation. To that end we use the ShengBTE[55] package, which offers a full-iterative solution of the Boltzmann transport equation to estimate the lattice thermal conductivity. The computationally demanding section of aforementioned approach is to acquire the third-order (anharmonic) interatomic force constants, which usually requires a few hundred or thousand single-point DFT calculations over supercell lattices. In this work, second and third-order force constants are calculated using the density functional perturbation theory simulations and passively trained MTPs, respectively, over 10×10×1 super-cells (consisting of 200 atoms). For the evaluation of the third-order anharmonic interatomic force constants, we consider interactions up to the eleventh nearest neighbours. In this case, by using the ShengBTE [55] package, we calculate the force constants using the MTP for 312 structures in a negligible time, which otherwise with DFT would require significant computational resources. On the basis of the MTP trained over AIMD simulations within the PBE/GGA functional, the diffusive lattice thermal conductivity of graphene is finally estimated to be 3600 W/mK, which falls within the experimentally measured values of 1500_5300 W/mK[29–32]. In Fig. 3b the cumulative lattice thermal conductivity of single-layer graphene as a function of mean free path using the MTP-based solution is compared with the existing full-DFT calculations, which show close trends. We note that depending on the type of exchange-correlation functional, as well as the chosen supercell size and cut-off distance, the obtained thermal conductivity of graphene varies substantially which explains the remarkable scattering in the available literature data. On the basis of PBE/GGA functional and using the ShengBTE package, the thermal conductivity of monolayer graphene at room temperature are predicted to be 1936[56], 3100[57], 3550[58], 3845[59], 3720[60], 3288[61] and 5500 W/mK[62]. In Table S2, a more elaborated comparison with different experimental and full-DFT theoretical works on the thermal conductivity of single-layer graphene is achieved, which confirms the accuracy of the accelerated approach in this work. In Fig. 3c the phonon dispersion relations of graphene predicted by the DFT- and MTP-based methods show a close agreement (see Section 1.5 of the Supplementary Information document for computational details). As it is well known, acoustic phonons are the main heat carriers in the graphene. Fig. 3d shows the good agreement for the contribution of different acoustic modes to the overall thermal conductivity of graphene using the MTP-based approach and full-DFT calculations[63–65]. In Fig. S3, the contribution of ZA, TA, LA and optical modes on the phonon's group velocity, scattering rate and Grüneisen parameter of single-layer graphene are also illustrated, and again show consistency with the existing full-DFT results. We further examine the validity of the proposed MTP-based method in predicting the lattice thermal conductivity, by considering the bulk silicon and InAs (see Section 7 of the Supplementary Information document for computational details). The acquired results shown in Fig. S4 reveal close agreement between the proposed MTP-based approach and existing experimental and full-DFT studies. Our results thus confirm that MTP potentials can be effectively used to estimate the lattice thermal conductivity, not only by classical NEMD simulations but also with the full-iterative solution of the Boltzmann transport equation. In the latter case, the MTP-based approach can yield accurate results but with substantially reduced computational cost of the evaluation of anharmonic interatomic force constants in comparison with commonly employed DFT-based solution, which is a highly promising finding.

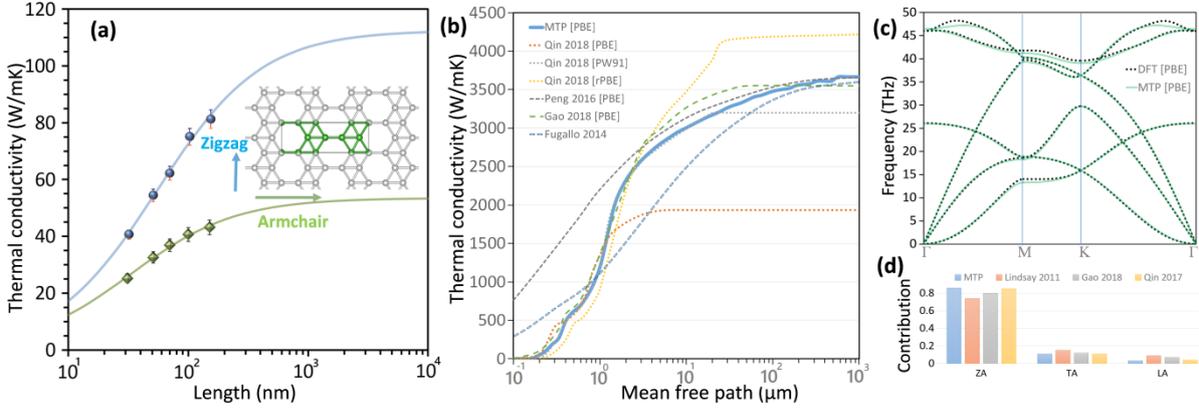

Fig. 3, (a) NEMD estimations for the length effect on the room temperature lattice thermal conductivity of single-layer borophene along the armchair and zigzag directions (continuous lines illustrate the fits to the NEMD data points). (b) Cumulative lattice thermal conductivity of graphene at the room temperature as a function of mean free path by fully iterative solutions of the Boltzmann transport equation using the MTP (present study) and full-DFT solutions by Fugallo et al.[66], Peng et al.[60], Gao et al.[64] and Qin et al.[56] with different exchange correlation functions. (c) Phonon dispersion relation of graphene acquired by the DFT (dotted line) and MTP (continuous line). (d) Contribution of ZA, TA and LA acoustic modes on the total lattice thermal conductivity of graphene by MTP (present study) and previous studies by Lindsay et al.[63], Gao et al.[64] and Qin et al.[65].

At this stage, we are capable to explore the effective lattice thermal conductivity of graphene/borophene heterostructures by employing the FEM simulations, in which we employ ABAQUS/Standard along with python scripting. For the construction of heterostructures, we develop polycrystalline samples made of 5000 individual grains on the basis of Voronoi cells with mirror symmetry at all edges[67]. Different grains are randomly assigned to be either graphene or borophene, according to the composition of heterostructures, simply by defining the corresponding thermal conductivity values acquired in the previous section. Since the borophene exhibits an anisotropic thermal transport, for the corresponding cells the anisotropic thermal conductivity tensors are defined by randomly selecting the orientation. The NEMD results for the thermal conductance of graphene/borophene grain boundaries are randomly chosen to introduce the interfacial conductance of every line connecting dissimilar crystals, and assuming perfect bonding (infinite conductance) for the rest of interfaces. To systematically investigate the size effect, we assume the equivalent grain size of the original polycrystalline sample as the domain size, assuming a square geometry for the equivalent average grain size[67]. A sample of heterostructure composition with 60% and 40% content of graphene and borophene phases, respectively, is shown in Fig. 4a. For the loading condition, we attach two highly conductive strips to the constructed sample and apply heat-fluxes ($h_f$) of the same magnitude on the outer surfaces of the two strips, one inward flux and one outward flux. As the initial value for the problem, the temperature of the outer surface of the cold strip (with outward flux) is set to zero. By solving the steady-state heat transfer problem, as shown in Fig. 4b a temperature profile establishes along the loading direction, which can be used to evaluate the effective thermal conductivity[67]. As expected and indeed observed in Fig. 4b-d, due to the high contrast in thermal conductivity of graphene and borophene, the temperature and heat flux profiles exhibit highly non-uniform distributions. This implies that the majority of heat fluxes are carried out via the percolation networks made of graphene crystals. To provide a more comprehensive vision on the heat transfer mechanism, we examine the effective lattice thermal conductivity for five different heterostructures and for domain sizes ranging from 1 nm to 100 μm (see Fig. 4e-h). It is clear that for all samples with extremely large domain sizes around 100 μm, the effective thermal conductivity is not yet fully converged, revealing the importance of assuming the interfacial thermal conductance for the modelling of thermal transport in these heterostructures. The presented results reveal three main behaviours of the effective thermal conductivity with respect to the domain size. Then first one occurring for domain sizes below 10 nm, the lattice thermal conductivity stays almost insensitive with respect to the domain size. This observation reveals that due to the presence of interfacial resistances, the embedded phases basically do not contribute to the heat flux transfer and exhibit a void like behaviour. This issue is noticeable when comparing the thermal conductivity of heterostructures with 10% and 20% content of graphene nanosheets (find Fig. 4h), in which the sample with the higher content of the ultrahigh conductive crystals yields lower conductivity for domain sizes lower than 100 nm. The second type of behaviour occurs for domain sizes from 10 nm to 10 μm, in which the thermal conductivity slowly increases with domain size. Such a trend implies that the effect of interfacial resistance starts to decline by increasing the domain size. If one considers for instance the sample with 10% content of borophene with a relatively large domain size of 250 nm, as seen in Fig. 4c, the borophene crystal contributes marginally to the heat flux transfer. For the sample with 40% content of borophene, it is noticeable that the majority of heat flux is transferred by graphene networks percolating each other (find Fig. 4d). In this case, borophene crystals not only participate marginally in the heat transfer but also impede thermal transport within the highly conductive graphene grains. In the third and last step, which dominates the thermal transport for domain sizes larger than 10 μm, the thermal conductivity reaches a plateau and only slightly increases by a further increase of the domain size, which reveals that the effect of interfacial resistance starts to vanish. From a practical point of view, this second step of heat transfer would be more close to real experimental samples, because it is normally very difficult to make heterostructures with domain sizes larger than 0.01 cm. Our results for domain sizes from 10 nm to 10 μm highlight that within these domain sizes, the interfacial thermal resistance plays the critical role and therefore should be taken into consideration. We would like to also clearly remind that in this study we mainly studied the lattice thermal conductivity, which may not be exactly the same as the total thermal conductivity, in which electrons contributions to the thermal conductivity are also taken into account.

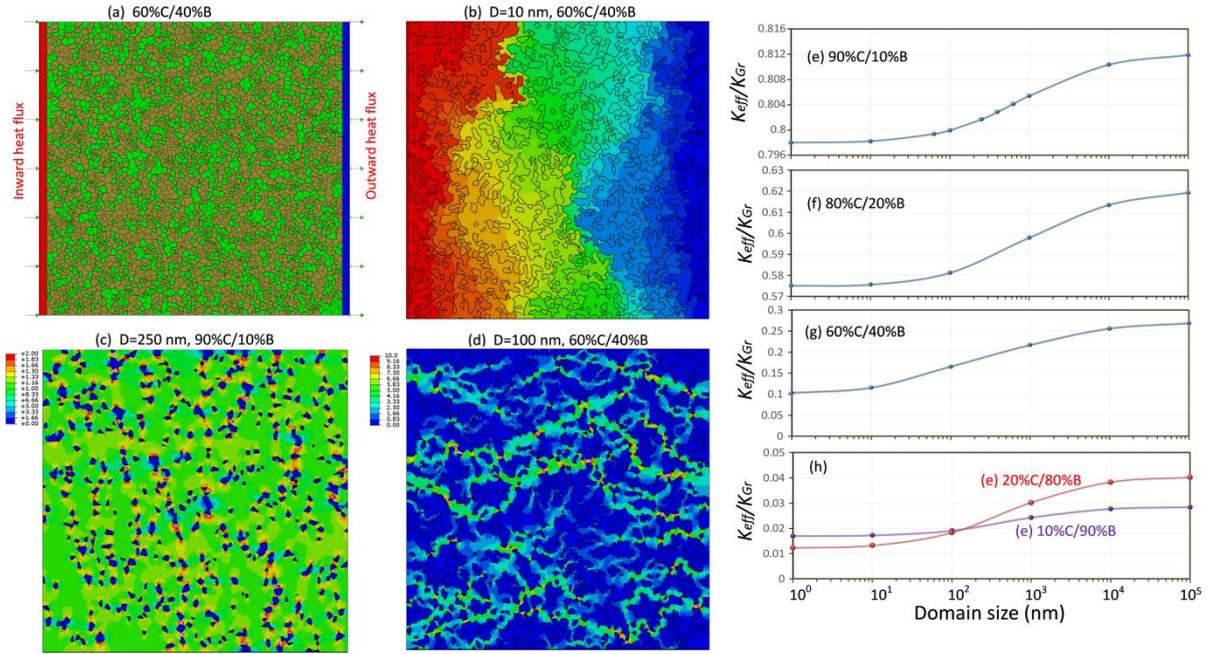

Fig. 4, (a) A samples of constructed continuum model of graphene/borophene heterostructure with 40% content of borophene crystals to evaluate the effective lattice thermal conductivity of polycrystalline graphene structures, (b) established steady-state temperature profile for the same sample with domain sizes of 10 nm. (c and d) Samples of heat flux distributions. (e to h) Normalized effective lattice thermal conductivity of heterostructure with respect to the graphene's thermal conductivity.

## Conclusion

In conclusion, our study confirms that machine-learning interatomic potentials trained over short ab-initio molecular dynamics trajectories enable efficient first-principles multiscale modeling via hierarchical employment of density functional theory/classical molecular dynamics/finite element simulations. In other words, it is possible to examine the properties and responses of novel complex microstructures, without prior knowledge of properties of building blocks. To show this novel possibility, we explored the lattice thermal conductivity of graphene/borophene heterostructures, a system that to the best of our knowledge there exist no viable classical modeling alternative. Furthermore, it is shown that the developed machine-learning interatomic potentials can be effectively employed to acquire the lattice thermal conductivity not only by classical molecular dynamics simulations but also with the full-iterative solution of the Boltzmann transport equation. First-principles multiscale modeling is believed to offer novel and computationally efficient possibilities to evaluate the properties and improve the design of advanced nanostructured materials.

## Methods

First-principles DFT calculations in this work were carried out using the *Vienna Ab-initio Simulation Package* (VASP)[68–70]. The generalized gradient approximation (GGA) and Perdew–Burke–Ernzerhof (PBE)[71] functional was adopted in the calculations. We assumed a plane-wave cutoff energy of 500 eV in our simulations. The phonon dispersion and second-order force constants of graphene were obtained by density functional perturbation theory (DFPT) simulations over a 10×10×1 supercell sample using a 3×3×1 Monkhorst-Pack[72] k-point grid along with the PHONOPY code[73]. Ab-initio molecular dynamics (AIMD) simulations were performed with a time step of 1 fs using a 3×3×1 k-point gird. For elaborated computational details, please refer to the supporting information document and the public Mendeley dataset of http://dx.doi.org/10.17632/pbgscy3ptg.1 .

Acknowledgment

B.M. and X.Z. appreciate the funding by the Deutsche Forschungsgemeinschaft (DFG, German Research Foundation) under Germany's Excellence Strategy within the Cluster of Excellence PhoenixD (EXC 2122, Project ID 390833453). E.V.P and A.V.S. were supported by the Russian Science Foundation (Grant No 18-13-00479). ICN2 is supported by the Severo Ochoa program from Spanish MINECO (Grant No. SEV-2017-0706) and funded by the CERCA Programme/Generalitat de Catalunya.

Data availability

The Supplementary Information document with computational details is available to download at http://dx.doi.org/10.17632/pbgscy3ptg.1. It contains the details of the computations, geometry optimized graphene/borophene heterostructures, examples of untrained MTPs, examples of VASP input parameters for the AIMD simulations, an example of non-equilibrium molecular dynamics simulation of thermal conductance with the MTP to introduce the atomic interactions using the LAMMPS package and developed codes to employ the MTPs for acquiring the anharmonic force constants for the BTE solution using the ShengBTE package, and the created input files for the ShengBTE solution of the lattice thermal conductivity of graphene, silicon and InAs.

# Supporting information

# Machine Learning Interatomic Potentials Enable First-Principles Multiscale Modeling of Lattice Thermal Conductivity in Graphene/Borophene Heterostructures

Bohayra Mortazavi*[a,b], Evgeny V. Podryabinkin[c], Stephan Roche[d,f],

Timon Rabczuk[g], Xiaoying Zhuang[a,g] and Alexander V. Shapeev[c]

[a]Chair of Computational Science and Simulation Technology, Department of Mathematics and Physics, Leibniz Universität Hannover, Appelstraße 11,30157 Hannover, Germany.
[b]Cluster of Excellence PhoenixD (Photonics, Optics, and Engineering–Innovation Across Disciplines), Gottfried Wilhelm Leibniz Universität Hannover, Hannover, Germany.
[c]Skolkovo Institute of Science and Technology, Skolkovo Innovation Center,
Nobel St. 3, Moscow 143026, Russia.
[d]Catalan Institute of Nanoscience and Nanotechnology (ICN2), CSIC and BIST, Campus UAB,
Bellaterra, 08193 Barcelona, Spain.
[f]ICREA Institució Catalana de Recerca i Estudis Avancats, 08010 Barcelona, Spain
[g]College of Civil Engineering, Department of Geotechnical Engineering,
Tongji University, Shanghai, China.

*bohayra.mortazavi@gmail.com

Related data are availbale via: http://dx.doi.org/10.17632/pbgscy3ptg.1

Content:
1. Training a moment tensor potential (MTP).
2. Graphene/borophene grain boundaries.
3. NEMD simulation using the MTP.
4. Phonon dispersion of graphene and borophene.
5. MTP/ShengBTE interface.
6. MTP/BTE results for graphene.
7. MTP/BTE results for bulk silicon and InAs.

# 1. Training a moment tensor potential (MTP).

## 1.1 Access to the MLIP package.

MLIP is a software package implementing MTP. It is distributed upon sending a reasonable request to Alexander Shapeev at *a.shapeev@skoltech.ru*. Please note that for the classical molecular dynamics simulations using the MTPs, the related plugin with LAMMPS [1] has to be also installed.

## 1.2 Creating training sets.

As explained in the original manuscript, the training sets are created by conducting the ab-initio molecular dynamics (AIMD) simulations at different temperatures. In our work, we employed the *Vienna Ab-initio Simulation Package* (VASP)[2–4] with generalized gradient approximation (GGA) and Perdew–Burke–Ernzerhof (PBE)[5]. In the Mendeley dataset, the folder entitled "*AIMD-inputs*", two samples of VASP input files (namely, POSCAR, POTCAR, INCAR and KPOINTS) for pristine borophene and a graphene/borophene heterostructure at 300 K are included. After the completion of AIMD simulations, OUTCAR file can be used to create the training set (*train.cfg*) with the following command:

*./mlp convert-cfg OUTCAR train.cfg --input-format=vasp-outcar*

Using the aforementioned command all the configurations will be included in the training set. Shortening of the training set (creating subsamples) can be achieved using the following command:

*./mlp subsample train.cfg subsample.cfg 10*

In the mentioned case, each first of every 10 snapshots in the original "*train.cfg*" will be written to the "*subsample.cfg*". The subsampled training sets at different temperatures or structures should be then merged together to create the final training set, which can be achieved using the Linux *"Cat Command"*.

## 1.3 Training of MTPs.

Training of MTP can be achieved by solving the following minimization problem:

$$\sum_{k=1}^{K}\left[w_e\left(E_k^{\text{AIMD}}-E_k^{\text{MTP}}\right)^2 + w_f\sum_{i}^{N}\left|f_{k,i}^{\text{AIMD}}-f_{k,i}^{\text{MTP}}\right|^2 + w_s\sum_{i,j=1}^{3}\left|\sigma_{k,ij}^{\text{AIMD}}-\sigma_{k,ij}^{\text{MTP}}\right|^2\right] \to \min,$$

where $E_k^{\text{AIMD}}$, $f_{k,i}^{\text{AIMD}}$ and $\sigma_{k,ij}^{\text{AIMD}}$ are the energy, atomic forces and stresses in the training set, respectively, and $E_k^{\text{MTP}}$, $f_{k,i}^{\text{MTP}}$ and $\sigma_{k,ij}^{\text{MTP}}$ are the corresponding values calculated with the MTP, *K* is the number of the configurations in the training set, *N* is the number of atoms in every and $w_e$, $w_f$ and $w_s$ are the non-negative weights that express the importance of energies and forces and stresses in the optimization problem, respectively, which in our study were set to 1, 0.1 and 0.001, respectively. We note that the weights for the energy and force are the default values. As an example, the training of a MTP can be achieved using the following command:

```
mpirun -n n_cores ./mlp train p.mtp train.cfg --energy-weight=1 --force-weight=0.1 --stress-weight=0.001 --max-iter=2000 --curr-pot-name=p.mtp --trained-pot-name=p.mtp
```

In the mentioned case, where "*n_cores*" is the number of cores used for parallel training of MTP, *"p.mtp"* is the input/output (curr-pot-name/trained-pot-name) MTP file, "*train.cfg*" is the training set in internal *.cfg MLIP format, the option "*max-iter*" determines the maximum number of iterations in the optimization algorithm. The options "*energy-weight*", "*force-weight*", and "*stress-weight*", respectively, define the $w_e$, $w_f$ and $w_s$ weights explained earlier.

*Important note:*

As discussed in the preparation of training sets, from the complete sets of AIMD configurations, only subsamples are selected for the training of first MTPs. Nonetheless, some critical configurations that could result in the improved stability of trained MTPs may have been missed in the created subsamples. Therefore, the accuracy of the developed MTP "*p.mtp*" using the initial subsampled training set "*train.cfg*" should be once again checked over the full AIMD configurations "*trainF.cfg*", and the configurations with high extrapolations grades [6] should be selected, and written to the file "*trainN.cfg*", via the following command:

*./mlp select-add p.mtp train.cfg trainF.cfg trainN.cfg*

The selected configurations "*trainN.cfg*" should be added to the original training sets "*train.cfg*" and the final MTP will be developed by retraining of new clean potentials over the updated training set. This way, the efficient use of conducted AIMD simulations will be guaranteed.

1.4 Structure of MTPs.

MTP belongs to the family of machine-learning interatomic potentials by which potentials show flexible functional form that allows for systematically increasing of the accuracy with an increase in the number of parameters and the size of the training. In the folder entitled "Untrained-MTPs", we included three samples of clean MTPs. Depending on the number of parameters, the appropriate MTP should be chosen. Prior to training there are few parameters to be adjusted, such as the "*species_count*", "*min_dist*" and "*max_dist*" which, respectively, define the number of element in the system, minimum atomic distance and cutoff distance of the potential. Please note that "*species_count*"and "*min_dist*" will be updated after the training process. Like the classical potentials, by increasing the cutoff distance more neighbors will be included in the calculations which will increase the computational costs accordingly. The number of parameters in a MTP can be calculated via:

*species_count$^2$.radial_basis_size.radial_funcs_count+alpha_scalar_moments+1*

Please note that *"radial_funcs_count"* and "*alpha_scalar_moments*" are the fixed features of a particular MTP and only "*radial_basis_size*" can be manually changed to adjust the number of constants.

### 1.5 Evaluation of phononic properties using the MTPs.

In our previous work, we included the full details and numerous examples for the evaluation of phononic properties using the MTP and PHONOPY [7] package in a public Mendeley dataset, please refer to: http://dx.doi.org/10.17632/7ppcf7cs27.1

## 2. Graphene/borophene grain boundaries.

Because of the different atomic structures of borophene and graphene and depending on the various tilting angles of intersecting crystals, graphene/borophene grain boundaries can show diverse configurations. In this work, as shown in Fig. 2a we constructed 10 different grain boundaries. Since plane-wave AIMD calculations were conducted, in order to create the training sets, the constructed models are ought to be periodic in planar directions. This way, in every graphene/borophene heterostructure model, two different grain boundaries were formed. In this work, we have chosen 6 graphene and borophene heterostructure models, and from those we have chosen 10 different grain boundaries for classical NEMD simulations. In the following illustration, these 6 models are illustrated. Constructed graphene/borophene heterostructure in the VASP native POSCAR format are included in the "*Graphene-Borophene-Heterostructures*" folder of the dataset:

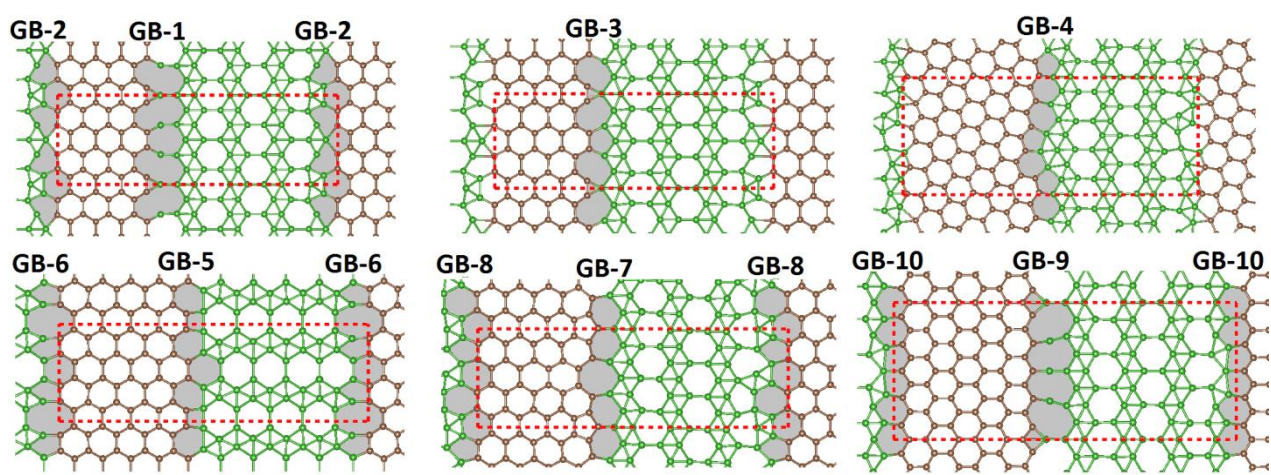

**Fig. S1**, Constructed graphene/borophene heterostructures for creating the training data. The dashed red-lines exhibit the boundary of models.

We examined the accuracy of trained MTPs for the constructed graphene/borophene heterostructures by conducting the AIMD simulations at 300 K for additional 4000 time steps. By including the original 1000 AIMD trajectories at 300 K, the final trajectories for every model include 5000 configurations. In the following table, the errors of the originally trained MTPs over 5000 AIMD configurations are summarized. Notably, as it can be seen from the results shown Table S1, the average absolute errors in different models are in less than 4 meV, which confirms the high accuracy of developed MTPs.

Table S1, Calculated errors of developed MTPs for 5000 AIMD configurations at 300 K.

|  | Graphene/borophene heterostructures models | | | | | |
| --- | --- | --- | --- | --- | --- | --- |
|  | GB1 & GB2 | GB3 | GB4 | GB5 & GB6 | GB7 & GB8 | GB9 & GB10 |
| C and B atoms, respectively | 32 and 48 | 32 and 45 | 56 and 61 | 40 and 40 | 37 and 45 | 60 and 64 |
| Average absolute difference in energy/atom (meV) | 3.4 | 2.2 | 1.4 | 1.5 | 2.0 | 1.0 |
| RMS of absolute difference in energy/atom (meV) | 4.0 | 2.6 | 1.8 | 1.9 | 2.7 | 1.3 |

## 3. NEMD simulation using the MTP.

In the provided Mendeley dataset, we included a sample of developed NEMD models for the calculation of a graphene/borophene grain boundary thermal conductance (find "*NEMD-Example*" folder). In order to define the interatomic potential type, in the LAMMPS ("*in.thermal*" in our example) script one has to use the following commands:

*pair_style mlip mlip.ini*
*pair_coeff * * *

in this case, "*mlip.ini*" is the interface with MLIP package, which includes the path to the trained MTP potential file (find "*mlip:load-from  p.mtp*"). The type of atoms in a MTP starts from 0 whereas in LAMMPS starts from 1, such that atomic type of 1 in the MTP (or in the training set) matches to the atomic type of 2 in the LAMMPS script. In the NEMD calculations, after applying the temperature difference at two ends and complete equilibration of the system, the applied heat fluxes by the NVT method to the hot and cold reservoirs and established temperature gradient will be averaged and recorded, to calculate the thermal conductivity or thermal conductance at the interface. In the provided example, heat fluxes and averaged temperatures at every slab are written in the "*fnvt.txt*" and "*T-X.txt*", respectively.

## 4. Phonon dispersion of graphene and borophene

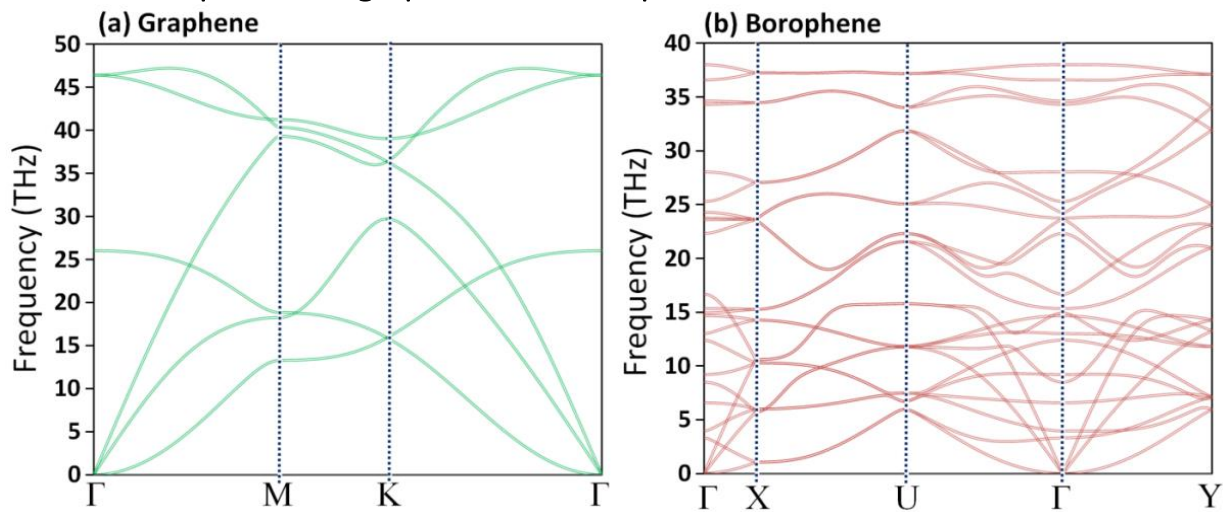

**Fig. S2**, Phonon dispersion relations of (a) graphene and (b) borophene. In this case the rectangular unit cell was considered for borophene.

## 5. MTP/ShengBTE interface.

ShengBTE [8] is a package for computing the lattice thermal conductivity on the basis of a full iterative solution to the Boltzmann transport equation. Its main inputs are sets of second- and third-order interatomic force constants and a CONTROL file for the adjustment of computational details. In this work, the calculation of anharmonic interatomic force constants is substantially accelerated by substituting DFT simulations with the MTP-based solution. For the calculation of anharmonic interatomic force constants, ShengBTE [8] provides a script, *"thirdorder.py"*, implementing a real-space supercell approach to anharmonic IFC calculations. In this approach, according to the defined supercell size and cutoff distance, the input geometries for the force constant calculations will be generated. For compatibility with *"cfg"*-file format, the *"thirdorder_vasp.py"* script is modified. Moreover, we developed an additional script *"fake_vasp_calcs.py"*, which uses the MTP-based calculated forces and artificially create the VASP output files of "vasprun.xml". This approach provides the possibility of direct comparison of forces by MTP and VASP.

In the folder "*MTP-ShengBTE-Examples*" the complete input files for the graphene, silicon and InAs are included. For every structure, the subfolder entitled "*ShengBTE-inputs*" includes the complete input files for the ShengBTE solution (namely: CONTROL, FORCE_CONSTANTS_2ND and FORCE_CONSTANTS_3RD). Using the data provided in the subfolder called "*Anharmonic-MTP*", a user will be able to reproduce the anharmonic interatomic force constants using the previously trained MTP instead of DFT simulations. To this aim, for every structure we included a shell script for complete calculations, named "*getFC.sh*". Please note that "*p.mtp*", "*mlip.ini*" and relate python scripts should be all in this folder for complete calculations.

## 6. MTP/BTE results for graphene.

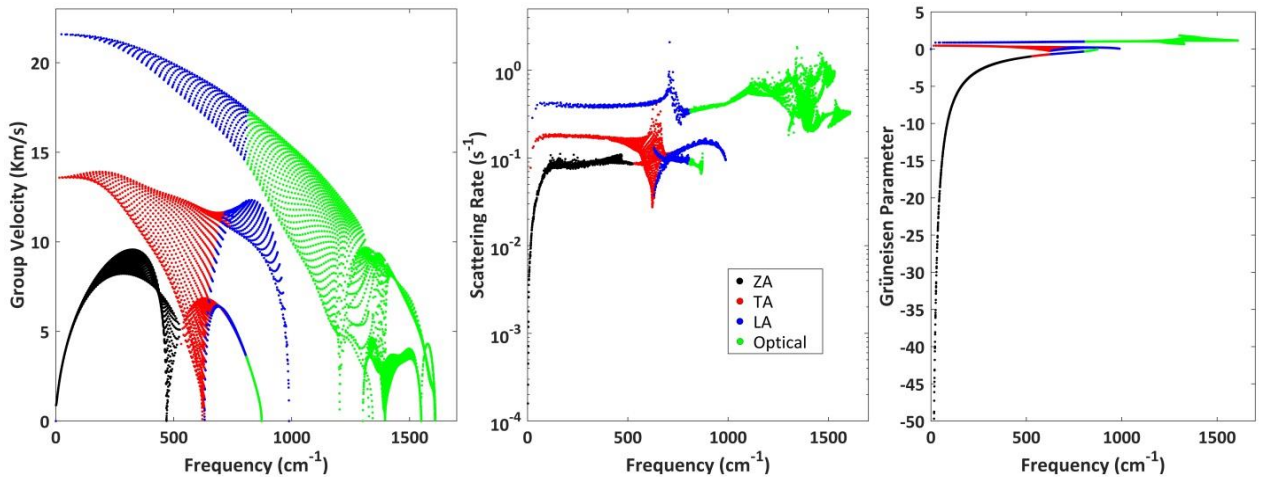

**Fig. S3**, Contribution of ZA, TA, LA and optical modes on the phonon group velocity, scattering rate and Grüneisen parameter of single-layer graphene as a function of frequency acquired on the basis of MTP/BTE solution.

Table S2, Comparison of thermal conductivity of single-layer graphene predicted using the MTP/BTE with different experimental and full-DFT/BTE theoretical reports.

| Thermal conductivity (W/mK) | Method |
|---|---|
| 3600 [300 K] | Present study, MTP/BTE |
| ~3080–5300 [300 K] | Raman spectroscopy [9,10] |
| 4127±539 [300 K] | Raman spectroscopy [11] |
| 1500-5000 [300 K] | Raman spectroscopy [12] |
| 2500±(+1100, -1050) [300 K] | Raman spectroscopy [13] |
| 2500 [300 K] | Micro electro-thermal (micro-resistance thermometer) [14] |
| 1689-1813 ±100 [300 K] | Micro electro-thermal (micro-resistance thermometer) [15] |
| 2430±190 [335 K] | Scanning thermal microscopy [16] |
| 632 [300 K] | Raman spectroscopy [17] |
| ~1800 [325 K] | Raman spectroscopy [18] |
| 2500 [310 K] | Raman spectroscopy [19] |
| 3100±1000 [350 K] | Raman spectroscopy [20] |
| 1937–2298 [300 K] | Micro electro-thermal (micro-resistance thermometer) [21] |
| 3720 [300 K] | DFT [PBE]/BTE [22] |
| 3550 [300 K] | DFT [PBE]/BTE [23] |
| 3000 [300 K] | DFT [PBE]/BTE [24] |
| 3600 [300 K] | DFT/BTE [25] |
| 3590 [300 K] | DFT [PBE]/BTE [26] |
| 3150 [300 K] | DFT [PBE]/BTE [27] |
| 1936 [300 K] | DFT [PBE]/BTE [28] |
| 5500 [300 K] | DFT [PBE]/BTE [29] |
| 3095 [300 K] | DFT [PBE]/BTE [30] |
| 3845 [300K] | DFT [PBE]/BTE [31] |
| 3288 [300 K] | DFT [PBE]/BTE [32] |

## 7. MTP/BTE results for bulk silicon and InAs.

To further examine the accuracy of proposed MTP/BTE approach in conjunction with the ShengBTE [8] software, we also study the thermal conductivity of bulk silicon and InAs, the examples considered in the original manuscript for ShengBTE [8]. We used VASP [2–4] package and PBE/GGA method with plane-wave cutoff energies of 330 and 300 eV for the silicon and InAs, respectively. The lattice constant of silicon and InAs were found to be 5.47 and 6.06 Å, respectively, which match closely with the those reported in Ref. [8]. The second-order force constants are obtained by DFPT simulations over 5×5×5 supercells using a 2×2×2 Monkhorst-Pack [33] k-point grid along with the PHONOPY code [7]. For the consistency with Ref. [8], anharmonic force constants are acquired by considering the interactions with the fourth nearest neighbours. Training sets are acquired by conducting the AIMD simulations at temperatures of 100, 300, 500 and 800 K for 1000 time steps. MTPs with 901 and 1009 parameters are trained for silicon and InAs, respectively. For the anharmonic force constants calculations, we also used 5×5×5 supercells. For the case of InAs, dielectric tensor and Born effective charges are also included in the calculations and the values were taken from the ShengBTE [8] examples. Fig. S4 compares the results by the MTP/BTE method with experimental and full-DFT/BTE solutions for the thermal conductivity of silicon and InAs. As it is clear, MTP/BTE results match closely with experimental and previous theoretical studies.

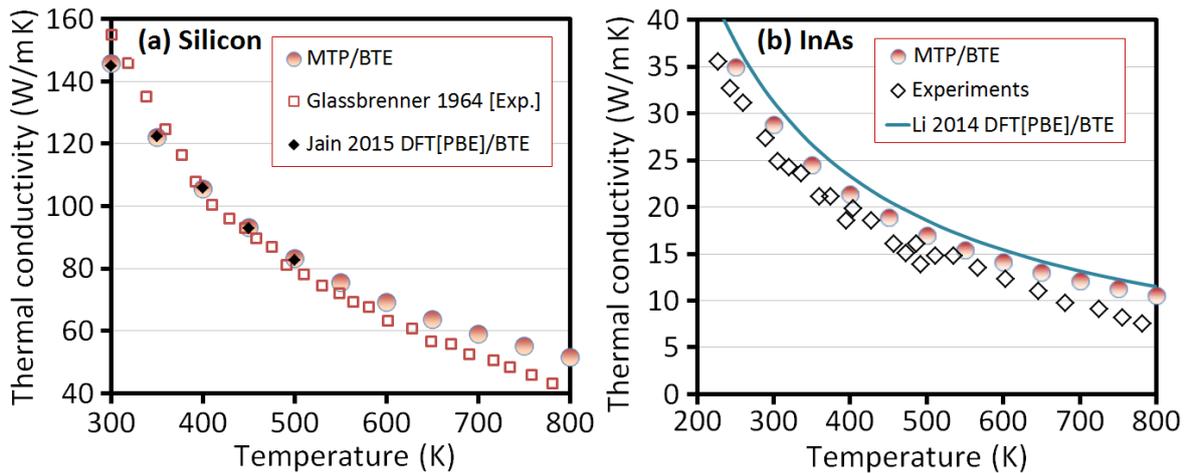

**Fig. S4**, Thermal conductivity of (a) silicon and (b) InAs predicted by the MTP/BTE and previous theoretical studies by Jain et al. [34] and Li et al. [8]. The experimental results for silicon and InAs were taken from [35] and [36,37], respectively.